\documentclass[%
 reprint,
 amsmath,amssymb,
 aps,
 floatfix
]{revtex4-1}

\usepackage{xcolor}
\usepackage{graphicx}
\usepackage[caption=false]{subfig}
\usepackage{dcolumn} 
\usepackage{bm} 
\usepackage{textcomp}

\graphicspath{{./images/}}

\renewcommand{\vec}[1]{\textbf{#1}}
\newcommand{\set}[1]{\left\{#1\right\}}
\newcommand{\dd}{\mathrm{d}}
\newcommand{\diff}[2][t]{\frac{\partial #2}{\partial #1}}

\begin{document}

\preprint{APS/123-QED}

\title{Entropy production and Vlasov equation for self-gravitating systems}

\author{Calvin A. Fracassi Farias}
\affiliation{Instituto de F\'isica, Universidade Federal do Rio Grande do Sul.\\}

\author{Renato Pakter}
\affiliation{Instituto de F\'isica, Universidade Federal do Rio Grande do Sul.\\}

\author{Yan Levin}
\email{levin@if.ufrgs.br}

\affiliation{Instituto de F\'isica, Universidade Federal do Rio Grande do Sul.\\}

\date{October, 2018}

\begin{abstract}
The evolution of a self-gravitating system to a non-equilibrium steady state occurs through a process of violent relaxation. In the thermodynamic limit the dynamics of a many body system should be governed by the Vlasov equation. Recently, however, a question was raised regarding the validity of Vlasov equation during the process or violent relaxation.  In this paper we will explore the entropy production during the relaxation process using N-body molecular dynamics simulations. We will show that the entropy production time grows as $N^\alpha$, with $\alpha>0$ and in the limit ${N \rightarrow \infty}$, entropy will remain constant, consistent with the Vlasov equation.  Furthermore, we will show that the mean field dynamics constructed on the basis of the Vlasov equation is in excellent agreement with the full molecular dynamics simulations, justifying the applicability of Vlasov equation during the violent relaxation phase of evolution.

\end{abstract}

\pacs{Valid PACS appear here} 

\maketitle

\section{\label{sec:introduction}Introduction}
Long range (LR) interacting systems  are distinct from  systems which interact through short-range forces. While the latter achieve thermodynamic equilibrium irrespective of the initial condition, the final state to which LR interacting systems evolve depends strongly on the initial condition. Self-gravitating systems (SGS) are paradigmatic of systems with LR interactions. It is known that SGS reach their  \textit{quasi}-stationary states (qSS) by process of violent relaxation \cite{LyndenBell1967, Campa2009, RibeiroTeixeira2014, Benetti2014}, in which some particles gain energy from the rest of the system through parametric resonances \cite{Benetti2014, Benetti2012, Levin2014, Gluckstern1994}. The process usually results in a violent relaxation to a qSS.   It has been well accepted that in the thermodynamic limit the dynamical evolution of the one-particle distribution function (DF) should be described by the Vlasov equation.  Recently, however, this belief has been questioned \cite{BeraldoeSilva2017} based on the investigation of the entropy production during the process of violent relaxation.
The authors of the reference \cite{BeraldoeSilva2017} observed for many different initial conditions a strong entropy increase during the process of violent relaxation which can not be accounted for in the framework of Vlasov equation, which requires that entropy must remain constant during the dynamical evolution.  

\begin{figure}[t!]
	\centering
    \includegraphics[width=\linewidth]{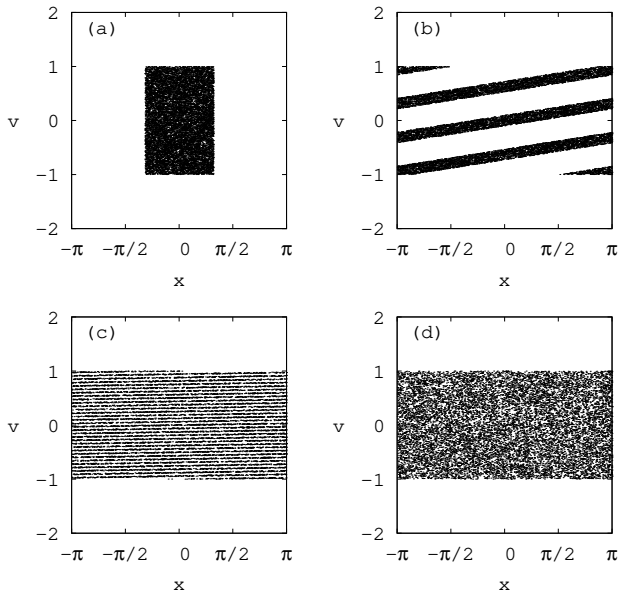}
    \caption{Snapshots of the phase space of a one dimensional system of  ${N = 131072}$ non-interacting particles with periodic boundary conditions: the fine-grained probability DF evolves through the process of filamentation  and phase-mixing. At some point the resolution is no longer sufficient to perceive  dynamics, and the system appears to be stationary. The  times of plots are: ${\text{(a) }t = 0}$, ${\text{(b) }t = 10}$, ${\text{(c) }t = 100}$ and ${\text{(d) }t = 1000}$.}
	\label{fig:_1dnip_snapshot}
\end{figure}

\begin{figure}[h!]
	\centering
    \includegraphics[width=\linewidth]{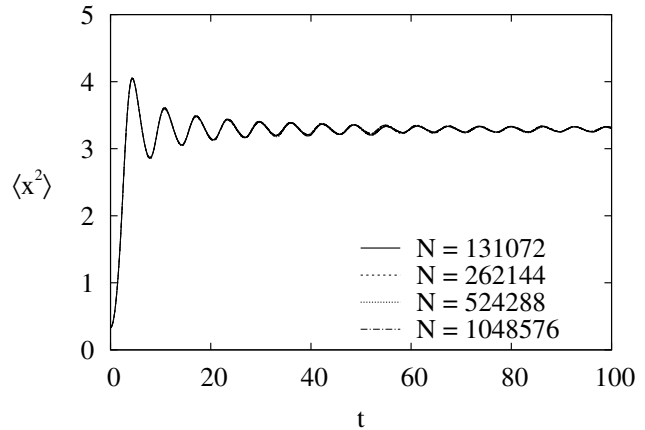}
    \caption{Violent relaxation of the second moment of the particle distribution, from initial to final stationary state.  Note that for sufficiently large system sizes, the relaxation time is independent of the number of particles in the system.}
	\label{fig:_1dnip_2ndM}
\end{figure}

For a $d$-dimensional system of particles interacting through a LR force, most of the contribution to the force acting on a given particle comes from the interaction with distant particles. In the thermodynamic limit, when ${N \rightarrow \infty}$, the pairwise interaction with the nearby particles can be neglected and the total force acting on a particle  can be calculated using the mean-field potential.  The probability DF of a many particle system can then be written in terms of a product on one-particle DFs 
\begin{equation} \label{eq:separable_pdf}
	f^{(N)}(\set{\vec{w}}, t) = \prod_{i=1}^N f(\vec{w}_i, t)
\end{equation}
which satisfy the collisionless  Boltzmann, or Vlasov, equation \cite{Binney2008, Spohn1991, Braun1977, Gabrielli2010}, 
\begin{equation}
	\left( \diff{} + \frac{\vec{p}}{m} \cdot \nabla_\text{q} - \nabla_\text{q} \psi \cdot \nabla_\text{p} \right) f \left( \vec{q}, \vec{p}, t \right) = 0
    \label{eq:Vlasov}
\end{equation}
where $\vec{q}$ and $\vec{p}$ are respectively the generalized coordinate and momentum, \emph{m} is the particle's mass, $f$ is the one-particle DF, and $\psi \equiv \psi(\vec{q}, t)$ is the mean-field interaction potential. The advantage of working with LR systems is that the $2dN$-dimensional phase space of systems with short range interactions effectively collapses to a $2d$-dimensional
$\mu$ phase space which can be more easily visualized and studied.
Since Vlasov equation is time-reversible, its microscopic dynamics needs to be reconciled with the Clausius second law of thermodynamics. In fact, Boltzmann solved a similar problem for systems with short-range interacts by postulating that the entropy is a logarithm of the total number of  microstates compatible with a given macrostate, irrespective of whether these microstates can be reached from a given initial condition or not \cite{Lebowitz1993}. On the other hand, the Gibbs entropy,  
\begin{equation}
	S_G = - k_B \int f^{(N)} \left( \set{\vec{w}}, t\right) \ln f^{(N)} \left(\set{\vec{w}}, t \right) \dd^N \vec{w} 
\end{equation}
where the integral is performed over all the phase space and ${\dd^N \vec{w} \equiv \set{\dd\vec{w}_1,...,\dd \vec{w}_N}}$, is conserved by the Liouville/Vlasov dynamics.   Since the Liouville and Vlasov equations can be written as  $\dd f^{(N)} / \dd t = 0$, the probability DF evolves as an incompressible fluid over the phase space, and any local integral of DF is conserved by the flow \cite{Chavanis2005}. The evolution of the probability DF never stops, continuing on smaller and smaller length scales. Therefore, only on a coarse-grained scale it is possible to say that a system evolves to a stationary state and that the entropy ``increases" \cite{Tremaine1986, Levin2008, Pakter2017}. This behavior is illustrated in Figure \ref{fig:_1dnip_snapshot}, which shows the evolution of $\mu$ phase space of a one dimensional system of non-interacting particles confined in a box with periodic boundary conditions, starting from an initial waterbag distribution. Clearly for this non-interacting system there is no doubt of validity of Vlasov equation. The initial distribution is seen to evolve through a process of filamentation and phase space mixing. During the dynamics, the initial distribution stretches and folds over an extended volume of the $\mu$ phase space, (Figures \ref{fig:_1dnip_snapshot}b and \ref{fig:_1dnip_snapshot}c).  On a fine-grained scale the phase space  volume occupied by the particles remains constant.  On a  coarse-grained scale, however, it appears that the evolution reaches a stationary state (Figure \ref{fig:_1dnip_snapshot}d), in which phase space volume occupied by the particles is larger than the volume of the original distribution. In Figure \ref{fig:_1dnip_2ndM} we also show the ``violent relaxation" of the second moment of the particle distribution function as it evolves to equilibrium from $t=0$. Note that the violent relaxation time does not depend on the number of particles, for sufficiently large system sizes, as was also observed for systems of interacting particles \cite{Barr2006}. 

For LR systems the entropy can be rewritten in terms of one-particle distribution function, Eq. \ref{eq:separable_pdf}, and can be calculated using an entropy estimator \cite{Dobrushin1958, Kozachenko1987, Singh2003}
\begin{eqnarray} \label{eq:entropyEst}
    \bar{s}_G k_B^{-1} = \frac{1}{N} \sum_{i=1}^N \ln\left(N r_i^{2d} V_{2d}\right) + \gamma
\end{eqnarray} 
where $r_i$ is the distance in the $\mu$ phase space from particle $i$ to its nearest neighbor, $V_{2d}$ is the volume of a hyper-sphere of $2d$ dimensions, $\gamma$ is the Euler-Mascheroni constant, $k_B$ is the Boltzmann constant and $N$ the number of particles. The quantity $\bar{s}_G k_B^{-1}$ is, in fact, a coarse-grained estimator of Shannon entropy per particle, which is distinguished from Gibbs entropy by the constant $k_B$. This facilitates comparison with \cite{BeraldoeSilva2017}, which also estimates Shannon entropy.

\begin{figure}[t!]
	\centering \subfloat{
    	\includegraphics[width=0.9\linewidth]{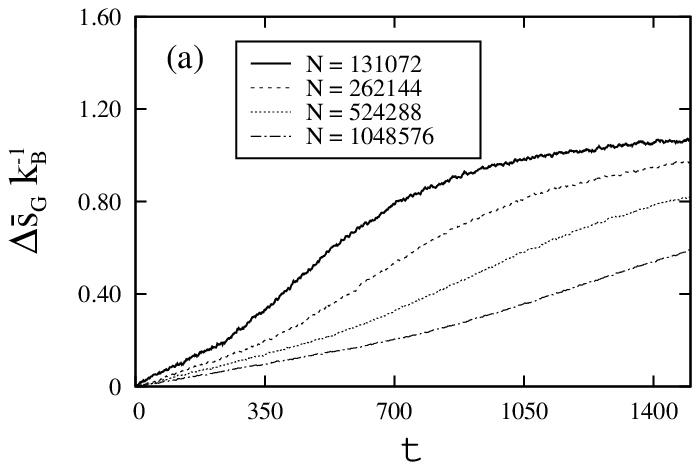}
        \label{fig:_1dnip_entropy}
    } \vfill
	\subfloat{
    	\includegraphics[width=0.9\linewidth]{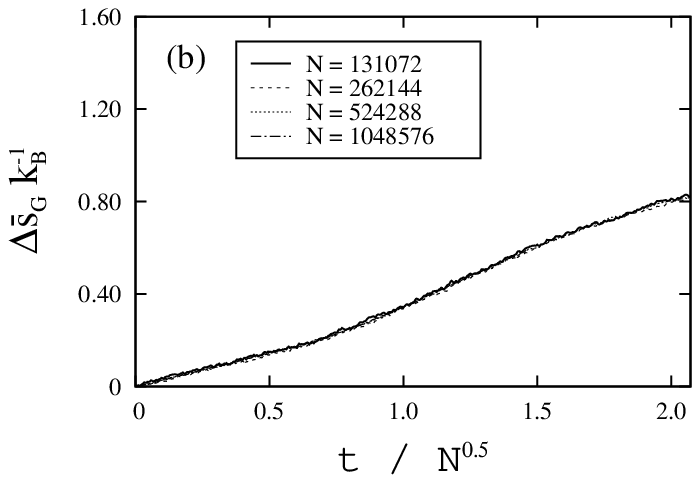}
        \label{fig:_1dnip_collapse}
    } \caption{(a) Entropy production per particle in a one dimensional system of non-interacting particles of Figure \ref{fig:_1dnip_snapshot}; and (b) is the data collapse. For this non-interacting system the collapse appears to be exact, showing that the entropy production time scales with $N^{\alpha}$, $\alpha = 0.5$. Therefore, in the limit ${N \rightarrow \infty}$, the entropy will remain constant, as is required by Vlasov equation.}
    \label{fig:1dnip}
\end{figure}

\begin{figure}[b!]
	\centering \subfloat{
    	\includegraphics[width=0.9\linewidth]{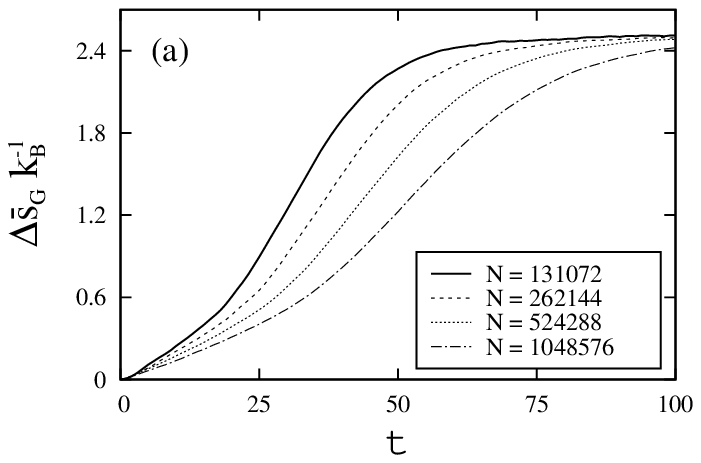}
        \label{fig:_2dnip_entropy}
    } \vfill
	\subfloat{
    	\includegraphics[width=0.9\linewidth]{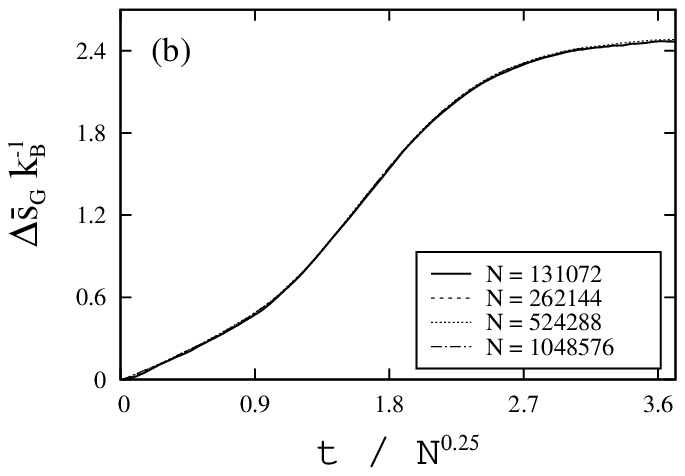}
\label{fig:_2dnip_collapse}
    } \caption{(a) Entropy production in a two dimensional system of non-interacting particles in a box with periodic boundary conditions; and (b) is the data collapse. Like Figure \ref{fig:1dnip}, the collapse appears to be exact, showing that the entropy production time scales with $N^{\alpha}$, $\alpha=1/4$. Therefore, in the limit ${N \rightarrow \infty}$, the entropy will remain constant, as is required by Vlasov equation.}
    \label{fi:2dnip}
\end{figure}

\begin{figure}[t!]
	\centering \subfloat{
    	\includegraphics[width=0.9\linewidth]{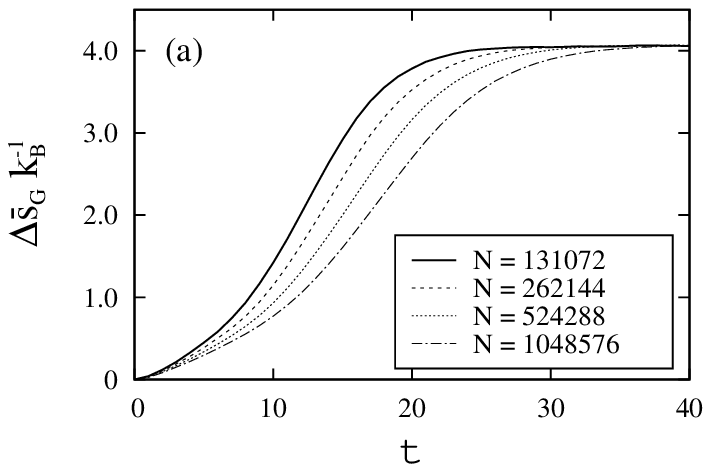}
        \label{fig:_3dnip_entropy}
    } \vfill
	\subfloat{
    	\includegraphics[width=0.9\linewidth]{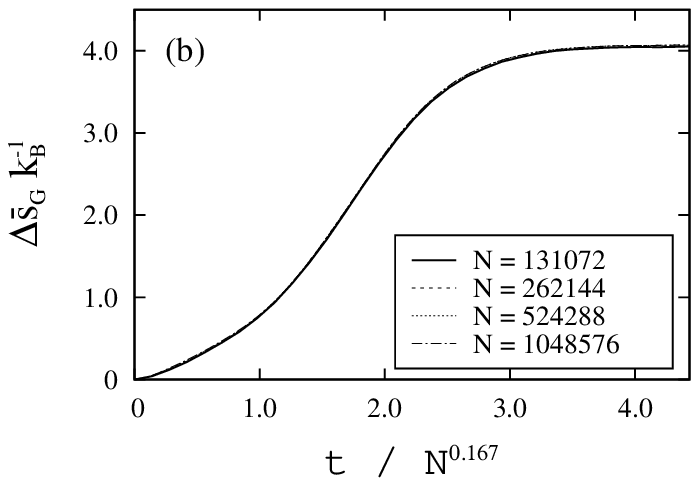}
        \label{fig:_3dnip_collapse}
    } \caption{(a) Entropy production for a three dimensional system of non-interacting particles inside a box with periodic boundary conditions; and (b) is the data collapse. If the entropy production time is scaled with $N^{\alpha}$, $\alpha = 1/6$ a perfect data collapse is observed.}
    \label{fig:3dnip}
\end{figure}

Figure \ref{fig:_1dnip_entropy} shows the entropy production (per particle) for a one dimensional non-interacting particle system of Fig. (\ref{fig:_1dnip_snapshot}) with various number of particles $N$. As expected, in spite of the system dynamics being governed by Vlasov equation, the coarse-grained entropy is not to conserved. On the other hand if the time is rescaled with $N^{1/2}$ we see that the entropy productions curves all collapse onto a universal curve. Figures \ref{fig:_2dnip_entropy} and \ref{fig:_3dnip_entropy} show the entropy production for a system of non-interacting particles in two and three dimensions, respectively. A perfect data collapse is again found, if time is rescaled with $N^\alpha$, where the exponent $\alpha$ is $\alpha=1/2d$.  Therefore, the entropy production time for non-interacting particles in $d$ dimensions scales as $\tau_\times \sim N^{1/2d}$, and diverges in the thermodynamic limit, implying that the fine-grained entropy will remain constant, as is required by the Vlasov equation. The fact that the coarse-grained entropy increases with time, does not invalidate in any sense Vlasov equation which is exact for these no-interacting systems.  Indeed, as we already saw in Figure \ref{fig:_1dnip_2ndM} a calculation of observables such as $\langle x^2 \rangle$ can be equally well performed using either a fine-grained distribution function or a coarse-grained one, in spite of the fact that coarse-grained entropy increases with time. 

In the remaining of this paper the entropy production of SGS will be investigated. The objective is to verify that in the thermodynamic limit, Vlasov equation does describe the dynamical evolution of a self-gravitating system, including the violent relaxation phase. The paper is organized as follow: Sec. \ref{sec:entropy_in_sgs} gives a brief review of one and two dimensional SGS; Sec. \ref{sec:_3dsgs_quantities} focuses on three dimensional SGS, the entropy production, and others observables; Sec. \ref{sec:conclusion} discuses the results and presents the conclusions.

\section{\label{sec:entropy_in_sgs}Entropy production in self-gravitating systems} 
The difficulty with studying three dimensional  self-gravitating systems is that they are intrinsically unstable.  Since the Newton gravitational potential is unbounded from below and is bounded from above, some particles can gain enough energy from the rest of the system to completely escape its gravitational attraction \cite{Levin2008, Padmanabhan1990}. This makes it very difficult to perform any kind of statistical analysis of the 3d gravitational clusters, except for very special virial initial conditions, studied in Section \ref{sec:_3dsgs_quantities} \cite{Levin2008}.  Therefore, most of our analysis will be performed using one and two dimensional self-gravitating systems, for which the gravitational potential is unbounded from above, preventing particle evaporation. The SGS MD simulations were performed in CUDA/C++ language, at constant energy, with rescaled  dimensionless variables, i.e., the equivalent of considering the system's total mass and the gravitational constant equals to unity. For one dimensional SGS, the numerical method applied was a fourth order implementation of the symmetric B3A method of Runge-Kutta-Nystroem with six stages from the C++ BOOST/odeint library \cite{Ahnert2011}. The error in energy was kept smaller than $2.0 \times 10^{-7} \,\%$. For two and three dimensional SGS, it was applied the CUDA algorithm of clustering tiles into thread blocks \cite{Nyland2007} with the improvement of loop unrolling. The numerical method was a fourth order symplectic integrator from \cite{Yoshida1990} and the error in the energy was kept smaller than $1.0 \times 10^{-3} \,\%$.

\subsection{Virial condition}
To study entropy production, all the MD simulations start with initial waterbag distribution which satisfies the virial condition, ${\cal R}_0=2 K/(2-d)U=1.0$, where $K$ is the initial kinetic energy, $U$ is the gravitational potential energy, and $d$ is the space dimension. Such distributions are expected to be as close to stationary as possible, without being an exact solution of Vlasov equation \cite{Braun1977, Teles2010}, and should rapidly relax to the 
qSS \cite{Yamaguchi2015}. This allows us to reduce the loss of resolution due to numerical imprecision resulting from strong oscillations and diminishes the system size necessary to observe the finite size scaling of the entropy production time. The initial particle distribution has the form,
\begin{equation}
f(\vec{q},\vec{p}) = \eta\, \Theta(\text{q}_M - |\vec{q}|)\, \Theta(\text{p}_M - |\vec{p}|)
\end{equation}
where $\text{q}_M$ and $\text{p}_M$ are, respectively, the boundary limits of coordinates and momenta in the $\mu$ phase space, $\eta$ is a constant of normalization whose value in 1d is $\eta = (4\text{q}_M \text{p}_M)^{-1}$; in 2d is  $\eta = (\pi \text{q}_M \text{p}_M)^{-2}$; in 3d  $\eta = (4 \pi/3)^{-2} (\text{q}_M \text{p}_M)^{-3}$,  and $\Theta$ is the Heaviside function. 
In this paper, the distances will be measured in units of $\text{q}_M$, mass of particles in the units of total mass $M$, and the time in the units of dynamical time $\tau_d$, so as to make equations of motion dimensionless. This is equivalent to setting ${\text q}_M = 1.0$, total mass to $M=1$, and the Newton gravitational constant to $G=1$. 
The energy per particle at time $t=0$ is then $\epsilon_0 = \frac{p_M^2}{6} + \frac{1}{3}$ in one dimension, $\epsilon_0 = \frac{p_M^2}{4} - \frac{1}{8}$ in two dimensions, and $\epsilon_0 = \frac{3 p_M^2}{10} - \frac{3}{5}$ in three dimensions,  and the virial condition reduces to $\text{p}_M=1$ for all $d$, see \cite{Levin2014, Pakter2013}.

\begin{figure}[b!]
    \centering
    \includegraphics[width=\linewidth]{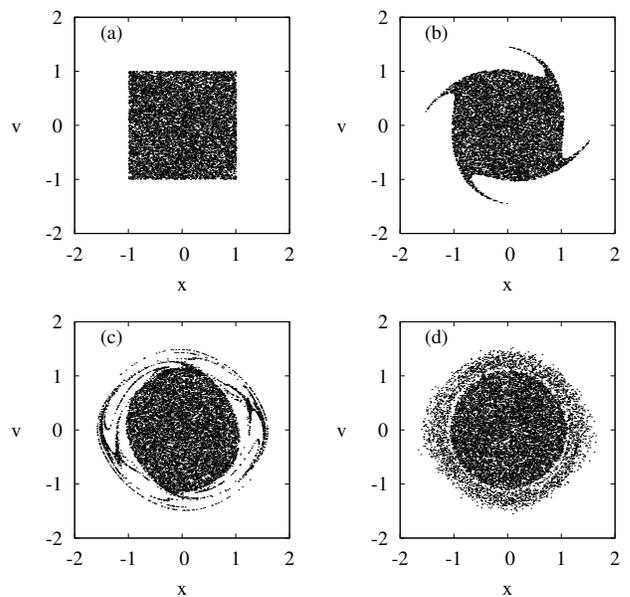}
    \caption{Snapshots of the phase space of a 1d self-gravitating system of section \ref{sec:1dsgs}. ${N = 131072}$ and ${R_0 = 1.0}$. The times are: ${\text{(a) }t = 0}$, ${\text{(b) } t = 10.0}$, ${\text{(c) }t = 100.0}$ and ${\text{(d) }t = 1000.0}$.}
    \label{fig:_1dsgs_snapshot}
\end{figure}

\begin{figure}[t!]
    \centering\subfloat{\label{fig:_1dsgs_entropy}
    	\includegraphics[width=0.9\linewidth]{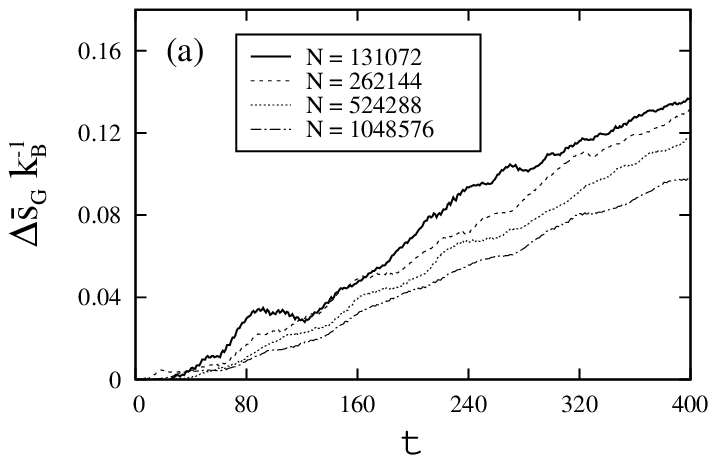}
    } \vfill
	\subfloat{\label{fig:_1dsgs_collapse}
    	\includegraphics[width=0.9\linewidth]{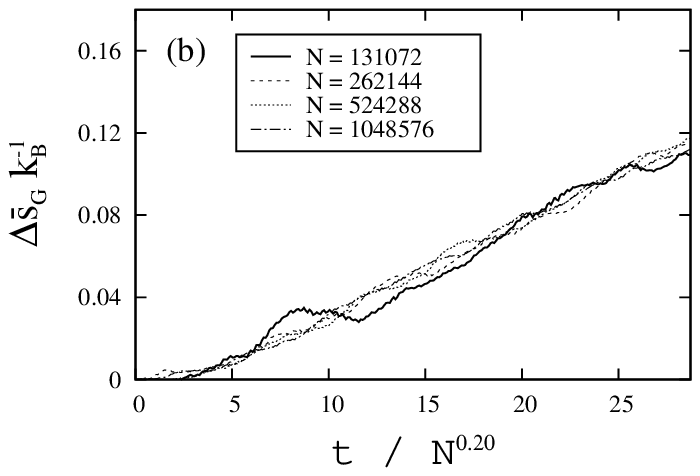}
    } \caption{(a) Entropy production for a 1d self-gravitating system of section \ref{sec:1dsgs}; and (b) is the time rescale. The early stages evolution indicate that the entropy production time grows as $N^{\alpha}$, $\alpha = 0.20$, therefore taking an infinite amount of time when $N \rightarrow \infty$. Different from non-interacting particles, scaling with $N$ appears to hold only at early times.}
\end{figure}

\subsection{\label{sec:1dsgs}Gravitation in one dimension}
One dimensional SGS consists of point particles of mass $m$ moving along the  $x$-axis, free to pass through one another. The reduced Poisson equation assumes the following form
\begin{equation}
	\nabla^2 \psi(x, t) = 2\,\rho(x, t)
    \label{eq:poisson1D}
\end{equation}
where $\psi(x,t)$ is the reduced gravitational potential and $\rho(x,t)$ is the reduced mass density. The reduced variables are: the dynamical time scale  $\tau_d = \sqrt{2\pi G\rho_0}$ where $G$ is the Newton gravitational constant; $\rho_0 = M / L_0$  is the mass density; $L_0$ is an arbitrary length scale which we take to be $q_M=1$; ${M = mN}$ is the system's total mass, which we set to 1; and $V_0 = \sqrt{2\pi G M L_0}$ is a velocity scale. The mass density of the i'th particle is $\rho(x, x_i) = m \delta(x - x_i)$. The reduced gravitational potential at position $x$ produced by $N$ particles is 
\begin{equation}
	\psi(x,t) = \frac{1}{N} \sum_{i}^N |x - x_i|
\end{equation}
Note that the binary interaction between any two particles vanishes as $1/N^2$. 
In the thermodynamic limit, therefore, the dynamics of a 1d SGS should be governed by the Vlasov equation.
The evolution of the particle distribution over the reduced $\mu$ phase space is shown in Figure \ref{fig:_1dsgs_snapshot}.
Once again we see the characteristic filamentation structure, which results in an effective gain of the phase space volume accessible to the particles, in the coarse-grained sense. The evolution of the coarse-grained entropy is shown in Figure \ref{fig:_1dsgs_entropy} for different system sizes. In  Figure \ref{fig:_1dsgs_collapse} we show that if the time is rescaled with $N^\alpha$, with $\alpha=0.2$, we can collapse the
entropy production onto a single curve.  This implies that in the thermodynamic limit $N \rightarrow \infty$, the time scale for the entropy production will diverge, and the entropy will remain constant consistent with the Vlasov dynamics.  This is similar to what was found for non-interacting particles, however, in the case of SGS the exponent $\alpha$ is smaller, implying that for systems with not too many particles the loss of fine-grained resolution happens very fast, leading to rapid entropy production. 

\begin{figure}[b!]
    \centering
    \includegraphics[width=\linewidth]{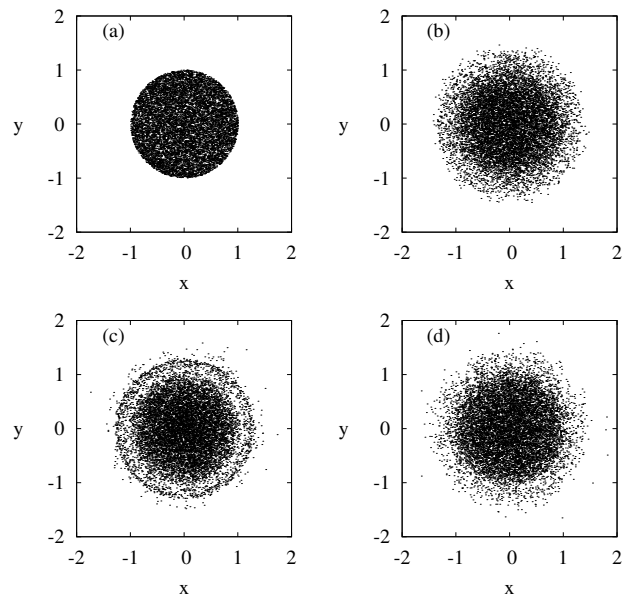}
    \caption{Snapshots of the configuration space of a 2d self-gravitating system of section \ref{sec:2dsgs}. ${N = 131072}$ and ${R_0 = 1.0}$. The  times are: ${\text{(a) }t = 0}$, ${\text{(b) } t = 10.0}$, ${\text{(c) }t = 32.0}$ and ${\text{(d) }t = 100.0}$.}
    \label{fig:_2dsgs_configurational_space}.
\end{figure}

\subsection{\label{sec:2dsgs}Gravitation in two dimensions}
\begin{figure}[t!]
    \centering\subfloat{\label{fig:_2dsgs_entropy}
       	\includegraphics[width=0.9\linewidth]{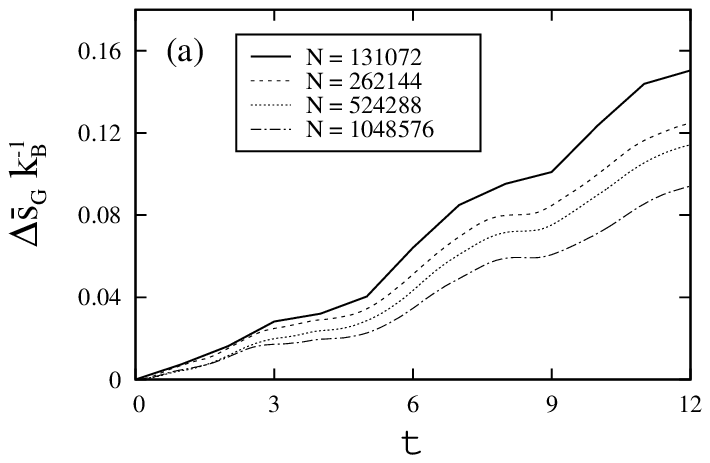}
    } \vfill
    \subfloat{\label{fig:_2dsgs_collapse}
    	\includegraphics[width=0.9\linewidth]{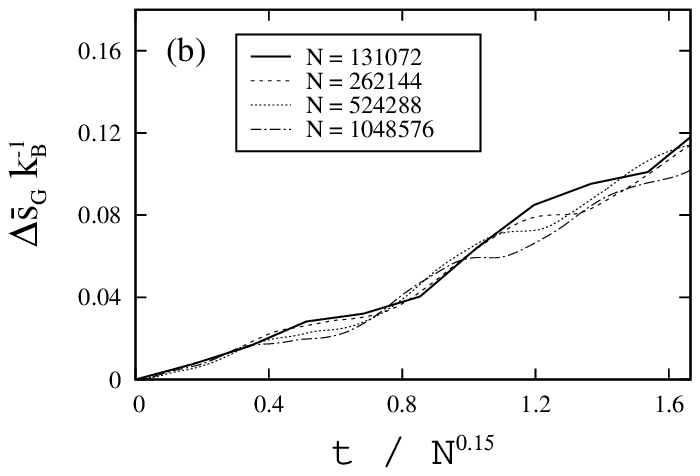}
    } \caption{(a) Entropy production in a 2d self-gravitating system with $N$ particles, section \ref{sec:2dsgs}; and (b) if the time is rescaled with $N^\alpha$, $\alpha = 0.15$, the curves can be reasonably collapsed to a single curve for short times.  In the thermodynamic limit, therefore, the entropy production will be zero, consistent with the Vlasov equation.}
\end{figure}

For a two dimensional SGS, the dimensionless Poisson equation is
\begin{equation}
	\nabla^2 \psi({\bf r}, t) = 2 \pi \, \rho({\bf r}, t)
    \label{eq:poisson2D}
\end{equation}
where $\psi({\bf r},t)$ is the two dimensional gravitational potential and $\rho({\bf r},t)$ is the mass density. The dynamical time scale is ${\tau_d = L_0 / \sqrt{2GM}}$, where once again, $G$ is the Newton gravitational constant; $M$ is the system's total mass, which we set to 1; and $L_0$ is an arbitrary length scale which we set to $q_M=1$. The mass density $\rho({\bf r},t)$ is obtained by integrating the probability DF over the momentum. For $N$ particle system the gravitational potential at position ${\bf r}$ is given by the solution of Poisson equation
\begin{equation}
	\psi({\bf r},t) = \frac{1}{N} \sum_{i}^N \ln|{\bf r} - {\bf r}_i|\,.
\end{equation}
The evolution of the configuration space is shown in Figures \ref{fig:_2dsgs_configurational_space} and of coarse-grained entropy and its rescaled form are shown in Figures \ref{fig:_2dsgs_entropy} and \ref{fig:_2dsgs_collapse}. Once again we obtain a reasonable data collapse for early times, with the exponent $\alpha=0.15$.

\subsection{\label{sec:entropy_growth}Entropy production in 3d}
\begin{figure}[t!]
    \centering
    \subfloat{\label{fig:_3dsgs_entropy}
    	\includegraphics[width=0.9\linewidth]{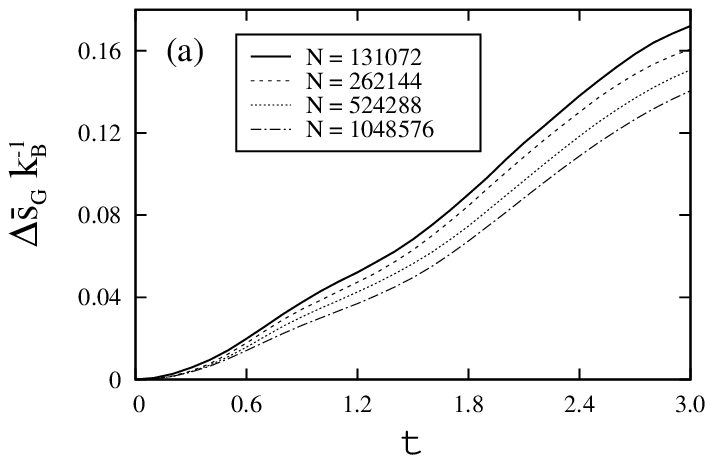}
    } \vfill
    \subfloat{\label{fig:_3dsgs_collapse}
    	\includegraphics[width=0.9\linewidth]{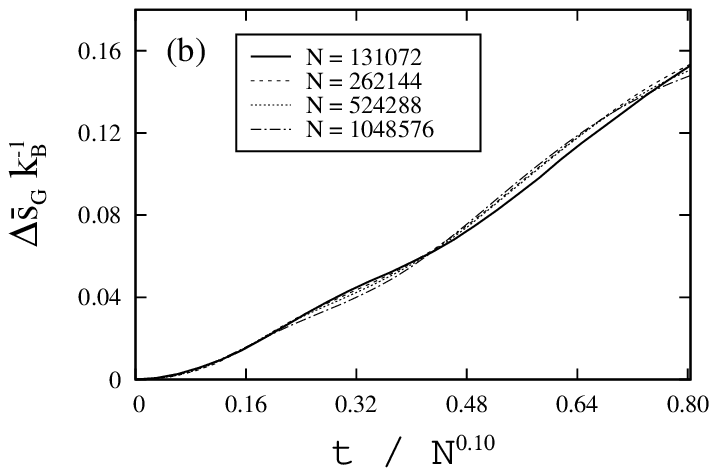}
    } \caption{(a) Entropy production in a three dimensional self-gravitating system and (b) its dynamical time rescale, $\alpha = 0.10$.
	}
\end{figure}

It is very difficult to study entropy production in 3d SGS because of a very rapid loss of resolution, which requires a very large number of particles to detect the scaling structure of the entropy production time. Nevertheless, in Fig. \ref{fig:_3dsgs_entropy} we see that if the dynamical time is rescaled by $N^{\alpha}$, the early time region of the entropy production curves collapses onto a single curve, showing that, at least in the early stages of the simulation, i.e., during the period of violent relaxation, there is a reasonably good scaling of entropy with the number of particles. The exponent $\alpha \approx 0.1$ for 3d systems, however, is lower than for 1d and 2d SGS. Nevertheless, in the limit $N \rightarrow \infty$ the entropy production will require infinite amount time, consistent with the Vlasov dynamics. 

\section{\label{sec:_3dsgs_quantities}3d SGS: evolution of observables}
In view of the very fast loss of resolution and rapid entropy production  in 3d SGS, 
the authors of Ref. \cite{BeraldoeSilva2017} argued that Vlasov equation is not appropriate to describe the violent relaxation of these systems.  Based on our finite size analysis, however, we see that this conclusion is incorrect, since in the infinite $N$ limit, the time for the entropy production diverges, and the fine-grained entropy will remain constant as is required by the Vlasov equation.  Nevertheless,
one might question whether for systems with large, but finite $N$, Vlasov dynamics can provide an accurate description of the temporal evolution of 3d SGS and, in particular their relaxation to the qSS.  Unfortunately, it is very difficult to explicitly solve the Vlasov equation for 3d SGS, however, we can explore the validity of the assumptions underlying Vlasov equation by performing simulations in which each particle interacts with the mean gravitational potential produced by all other masses. Such simulations will eliminates the correlational (or collisional) effects and provides an indirect way of solving the Vlasov equation.  We shall call such simulations ``collisionless MD".  For spherically symmetric particle distributions, the mean-field can be easily calculated by replacing each particle by a spinning spherical shell of radius and angular momentum same as the real particle. For the shell system, the force is purely radial, so that the angular momentum of each shell is conserved.  The dynamics of each shell then reduces to its radial coordinate, and the force on each shell can be easily calculated using the Gauss law~\cite{Levin2014}. Clearly if both collisional (simulation which is based on explicit binary interaction between the particles) and collisionless simulations will result in the same dynamical evolution of observables of a system, it will provide a very clear indication of validity of Vlasov dynamics for systems with large but finite number of particles, in spite of the entropy production. 

\begin{figure}[b!]
	\centering
    \includegraphics[width=\linewidth]{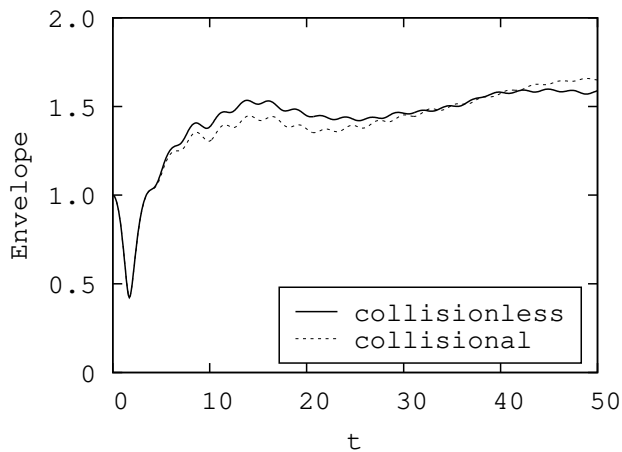}
    \caption{Envelope of the particle distribution. Both collisional and collisionless simulations follow identical dynamical evolution, implying that Vlasov equation accounts perfectly well for the violent relaxation phase, ${N = 131072}$ and ${R_0 = 0.5}$.}
    \label{fig:_3dsgs_envelope}
\end{figure}

\begin{figure}[t!]
	\centering
	\includegraphics[width=\linewidth]{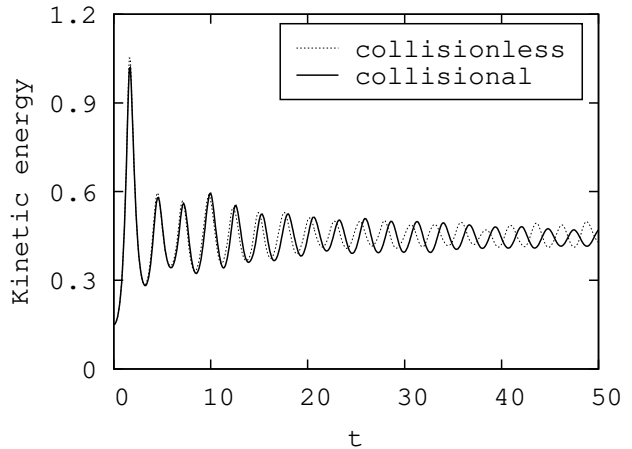}
    \caption{Kinetic energy per particle. Both collisional and collisionless simulations have almost identical dynamics. A small difference between the curves is due to slightly different initial conditions in the two simulations verified using different seeds for the pseudo-random number generator in the MD  simulations. $N=131072$ and $R_0 = 0.5$.}
    \label{fig:_3dsgs_kinetic}
\end{figure}

One particularly relevant quantity to study in MD simulations is the evolution of the ``envelope" of the particle distribution defined in terms of the root-mean-squared (rms)  of the particle coordinates \cite{Levin2014}
\begin{equation}
	r_e = \sqrt{\frac{5}{3} \left \langle \vec{r} \cdot \vec{r} \right \rangle}\,.
\end{equation}
The factor of $5/3$ is included so that at $t_0$, the envelope is precisely $\text{q}_M$. The other interesting quantity to consider is the average kinetic energy of the particles.  We will compare the evolution of both the envelope and the kinetic energy using both collisional and collisionless  MD simulations for initial distribution with virial number $R_0 = 0.5$ and a number of particles 
${N = 131072}$. This virial number was chosen to force the system to undergo strong oscillations, rapid entropy production, while preserving the spherical symmetry of the initial distribution \cite{Pakter2013}.  Figure \ref{fig:_3dsgs_envelope} shows the time evolution of the envelope, while Figure \ref{fig:_3dsgs_kinetic} shows the evolution of the kinetic energy per particle. We see that both
collisional and collisionless simulations lead to almost identical evolution of both observables, in spite of a rapid entropy production. A small deviation in the final qSS, is due to slightly different initial conditions, due to random number generator.

Therefore, we  conclude that for 3d SGS the relevant observables can be equally well calculated using either the exact fine-grained distribution $f({\bf p}, {\bf q}, t)$ obtained from the solution of the Vlasov equation, or an effective coarse-grained distribution in which the $f({\bf p}, {\bf q})$ is coarse-grained over some microscopic length scale,
\begin{equation}\label{fbar}
f_{cg}(\mathbf{q}, \mathbf{p}, t) = \frac{1}{(\Delta p \Delta q)^{d}} \int_{\Delta p, \Delta q} f(\mathbf{q'}, \mathbf{p'},t)\,\dd\mathbf{q'} \dd\mathbf{p'} \,.
\end{equation}
While the entropy calculated using the fine-grained distribution is strictly conserved, the entropy calculated using the coarse-grained distribution will grow and saturate as is observed in the collisional MD simulations.  The role of ``coarse-graining" in the MD simulations comes from the residual correlations between the masses and the numerical error.  Nevertheless, the agreement between collisionless and collisional simulations implies that the average of observables calculated using either fine-grained or coarse-grained distributions are identical in the thermodynamic limit,
\begin{eqnarray}
&&\langle O(t) \rangle = \int O(\mathbf{q}, \mathbf{p}) f(\mathbf{q}, \mathbf{p},t)\dd\mathbf{q} \dd\mathbf{p} = \nonumber \\
&& \int O(\mathbf{q}, \mathbf{p}) f_{cg}(\mathbf{q}, \mathbf{p},t)\dd\mathbf{q} \dd\mathbf{p}.
\end{eqnarray}
The fine-grained distribution function obtained from the solution of the Vlasov equation
can, therefore, be used to account for the violent relaxation in SGS with large but finite number of particles.

\section{\label{sec:conclusion}Conclusions} 

We have explored the entropy production in SGS in one, two, and three dimensions. We find that the entropy production time scales as $N^\alpha$, with $\alpha = 0.20$, $0.15$, and $0.1$, respectively. It is not clear what precisely determines the value of the exponent $\alpha$. It decreases with the increasing dimensionality of configuration space and may also be related to the Lyapunov spectrum \cite{Antunes2015}. The loss of resolution happens very fast for 3d SGS. Contrary to the suggestion of \cite{BeraldoeSilva2017}, this however does not imply a failure of Vlasov equation to describe the process of violent relaxation. Indeed Vlasov dynamics is entropy conserving. This, however, is only valid in the limit $N \rightarrow \infty$. For finite systems, therefore, there will be a rapid loss of resolution which can be associated with the entropy production. Indeed within the Vlasov formalism, we can define a coarse-graining procedure, associated with the loss of resolution, which will also result in the growth of entropy. Such coarse-graining is very similar to the Boltzmann definition of entropy which counts the total number of  microstates compatible with a given macrostate, irrespective of whether these microstates can be reached from a given initial condition or not.
 Comparing the collisionless and collisional MD simulations we saw that the dynamics of observables in systems with relatively small number of masses $N$ is equally well described by either of the two simulation methods.
The collisionless simulations are effectively a solution of the Vlasov equation, while collisional simulations include residual correlations which lead to the entropy production in a finite $N$ system, the equivalence of the two methods implies that the evolution of the observables in SGS can be equally well calculated either using exact fine-grained distribution function or using its coarse-grained version. The same conclusion was also reached by analyzing non-interacting particle systems for which Vlasov equation is exact. We thus conclude that conservation of fine-grained entropy by Vlasov equation does not invalidate it in any way from providing an accurate description of violent relaxation dynamics that leads to quasi stationary states with increased coarse-grained entropy.

\begin{acknowledgments}
This work was partially supported by the CNPq, National Institute of Science and Technology Complex Fluids INCT-FCx, and by the US-AFOSR under the grant 
FA9550-16-1-0280.
\end{acknowledgments}

\bibliographystyle{unsrt}
\bibliography{references}

\end{document}